\documentclass[copyright,creativecommons]{eptcs}
\usepackage{breakurl}        

\usepackage{graphicx}
\usepackage{amsmath}
\usepackage{amssymb}
\usepackage{amsthm}
\usepackage{proof} 
\usepackage{stmaryrd}
\usepackage{graphicx}
\usepackage{multirow}
\usepackage{marvosym}
\usepackage[mathscr]{euscript}
\usepackage{pstricks,pst-node,pst-text,pst-3d}
\usepackage[all,color,dvips]{xy}
\usepackage{upgreek}
\usepackage{bbm}
\usepackage{cmll}

\usepackage[varg]{pxfonts}
\usepackage{dsfont}

\newtheorem{definition}{Definition}
\newtheorem{proposition}{Proposition}
\newtheorem{theorem}{Theorem}

\def\tuple#1{\langle #1 \rangle}
\def\pi{\uppi}

\def\category#1{\mathbb{#1}}

\def\homset#1#2#3{\category{#1}(#2,#3)}
\def\comp{\circ}
\def\id#1{id_{#1}}
\def\morph#1{\stackrel{#1}{\rightarrow}}

\def\product{{\, \pmb\times \,}}

\def\eval{{\mathrm{eval}}}
\def\curry#1{{\mathrm{curry}(#1)}}

\def\unitObj{{\mathbf 1}}

\def\unitLLaw{\uplambda}
\def\unitRLaw{\varrho}
\def\symLaw{\gammaup}

\def\Var{\mathrm{Var}}

\def\PCF{\mbox{PCF}}
\def\num#1{{\tt \underline{#1}}}
\def\succ{{\tt succ}}
\def\pred{{\tt pred}}

\def\Sllambda{{\mathcal S} \ell \lambda}

\def\FV{\mbox{FV}}
\def\lin{\multimap}

\def\Var{\mathrm{Var}}
\newcommand{\sVar}{\mathrm{SVar}}
\newcommand{\hVar}{\mathrm{HVar}}

\newcommand{\Sl}{{\mathcal{S}\hspace{-.4mm}\ell}}
\def\redSl{\rightarrow_{\Sl}}
\def\PCF{\ensuremath{\mathrm{PCF}}}

\newcommand{\SlPCF}{\Sl\mathrm{PCF}}

\def\lif#1#2#3{\ell {\tt if} \; #1 \; #2 \; #3}
\def\lifz{\ell {\tt if}}

\def\den#1{\llbracket #1 \rrbracket}

\def\Coh{\mathbf{Coh}}
\def\tr{\mathrm{tr}}

\def \which{{\tt which}?\,}

\def\Cl{\mathcal{C}l}

\newcommand\COMMENT[1]{}

\author{Marco Gaboardi
\institute{Dipartimento di Informatica\\
Universit\`a degli Studi di Torino}
\email{gaboardi@di.unito.it }
\and
Mauro Piccolo
\institute{Dipartimento di Informatica\\
Universit\`a degli Studi di Torino}
\institute{Preuves, Programmes et Syst\`emes\\ Universit\'e de Paris VII }
\email{piccolo@di.unito.it }
}
\title{Categorical Models for a Semantically Linear $\lambda$-calculus\thanks{Work partially supported by MIUR-PRIN'07 CONCERTO Project.}}
\def\titlerunning{Categorical Models for a Semantically Linear $\lambda$-calculus}
\def\authorrunning{Marco Gaboardi and Mauro Piccolo}
\begin{document}
\maketitle
\begin{abstract}
This paper is about a categorical approach to model a very simple  Semantically Linear $\lambda$-calculus, named 
$\Sllambda$-calculus. This is a core calculus underlying the 
programming language $\SlPCF$. In 
particular, in this work, we introduce the notion of $\Sllambda$-Category, 
which is able to describe a very large class of sound models of 
$\Sllambda$-calculus. $\Sllambda$-Category extends in the natural way
Benton, Bierman, Hyland and de Paiva's Linear Category, in 
order to soundly interpret all the constructs of $\Sllambda$-calculus. This 
category is general enough to catch interesting models in Scott Domains and
Coherence Spaces.
\end{abstract}

\section{Introduction} 
$\Sllambda$-calculus - acronym for $\mathcal{S}$emantically $\ell$inear 
$\lambda$-calculus - is a simple term calculus based on $\lambda$-calculus. 
More specifically, $\Sllambda$-calculus extends and refines simply typed 
$\lambda$-calculus by imposing a linearity discipline on the usage of certain 
kinds of variables, as well as by adding some programming features to the 
calculus - like numerals, conditional and fix point operators - to make the 
calculus expressive enough to program all first-order computable functions.

Semantically Linear $\lambda$-calculus was already introduced in 
\cite{paolini08ppdp} (with an additional operator called $\which$, that is not
present here) as the term rewriting system on which the programming language
$\SlPCF$ is based  \cite{gaboardi07siena,paolini08ppdp}. $\SlPCF$ is 
based on a syntactical restriction of $\PCF$  conceived in order to 
program only linear functions between Coherence Spaces. 
In particular, in
\cite{paolini08ppdp} we define a \emph{concrete} model of $\SlPCF$ (and consequently of 
$\Sllambda$-calculus) in the category $\Coh$ of Coherence Spaces and Linear 
Functions, for which we prove a full abstraction result. 

The aim of this paper is to give an \emph{abstract} description of models of 
$\Sllambda$-calculus. This in order to highlight the properties 
that a mathematical structure must satisfy to model, by means of its equational theory, 
the operational theory induced by the reduction rules of the calculus.
We give this abstract description in terms of category theory and we show that the obtained notion can be 
used to build concrete models in different 
mathematical structures.


We recall that the category $\Coh$, as well as many other categories, is a
well known concrete instance of Benton, Bierman, Hyland and de Paiva's Linear 
Categories, introduced in \cite{benton92csl} to provide an abstract description
of models of Intuitionistic Linear Logic. All these categories are
symmetric monoidal closed and they are equipped with a symmetric monoidal 
comonad $!$ used to interpret the exponential modality and satisfying certain
properties \cite{benton92csl}. The idea is to impose enough conditions on the
comonad in order to make its induced Kleisli category a Cartesian Closed 
Category with exponential object $A \Rightarrow B = !A \lin B$.
The original construction does not require this, but it would 
actually be the case, if the monoidal closed category is also 
cartesian.

In this paper, we introduce the notion of $\Sllambda$-Category which extends 
in the natural way the definition of Linear Category, in order to be able to 
interpret all programming constructs of $\Sllambda$-calculus. 

We ask that this category admits a morphism acting like a ``conditional'' and a 
morphism acting like a ``fix-point operator''. The latter turns out to be the 
expected decomposition of a fix-point morphism in a Cartesian Closed Category.
Furthermore, to interpret ground values, we require the existence of
a distinguished object $N$ with the usual zero and successor and predecessor 
morphisms satisfying the expected equations. However, since variables of ground 
type can be freely duplicated and erased, we need to ask that all numeral 
morphisms behaves properly with respect to the comonad $!$. For this purpose 
we ask the existence of $!$-coalgebra $p:N \to !N$ which is also comonoidal and 
moreover we ask that all numeral morphisms are both coalgebraic and comonoidal.

The notion of natural number object in a symmetric monoidal closed category
is not new and it was introduced by Par\'e and Rom\'an in \cite{pare88sl}.
Based on this definition Mackie, Rom\'an and Abramsky introduced an internal
language for autonomous categories with natural number objects in 
\cite{mackie93acs}. The main similarity between the definitions of natural
number object given in \cite{pare88sl,mackie93acs} and our definition is the
requirement of comonoidality of the natural number object; moreover their 
definition does not take into consideration the relationship between the 
natural number object and the exponential comonad $!$; in fact there, only 
a strictly linear language without exponential was analyzed. More details on 
this matter can be found in \cite{piccolo09phd}.

We prove that the proposed categorical model of $\Sllambda$-calculus enjoys 
soundness with respect to the smallest equivalence containing  
$\Sllambda$-reduction. This Soundness Theorem relies on three distinct 
substitution lemmas corresponding to the three kind of substitution in the 
calculus.

Moreover, this abstract definition of model for $\Sllambda$-calculus allows 
us to analyze in a modular way many different concrete examples. 
In particular,
we build a non-trivial model of $\Sllambda$-calculus in the category 
$\mathbf{StrictBcdom}$ of Scott Domains and strict continuous functions. We also
study models of $\Sllambda$-calculus in the category $\Coh$ of Coherence Spaces
and linear stable functions and in the category $\mathbf{StrictBcdom}$
of Scott Domains and linear functions.
More specifically, this implies that the model we defined in 
\cite{paolini08ppdp} is equivalent to a particular instance of 
$\Sllambda$-Category, in the category $\Coh$.

Finally, we address the completeness of the $\Sllambda$-calculus with
respect to $\Sllambda$-category. We show that the completeness
with respect to the standard interpretation fails. So we discuss
some ways to recover it.

We conclude by giving some possible future directions.



\section{Semantically Linear $\lambda$-calculus}
$\Sllambda$-calculus is a term rewriting system very close to 
$\lambda$-calculus, on which the programming language $\SlPCF$ 
is based \cite{gaboardi07siena,paolini08ppdp}.
Truth-values of $\Sllambda$-calculus are encoded as integers (zero encodes 
``true'' while any other numeral stands for ``false''). 
The set of \emph{$\Sllambda$-types} is defined as,
$\sigma, \tau::= \iota \ |\ (\sigma\lin \tau)$
where $\iota$ is the only \emph{atomic} type (i.e. natural numbers), $\lin$ is the only {\em type constructor} and $\sigma, \tau, ...$ are meta-variables ranging over types.
Let $\Var^{\sigma},\sVar^{\sigma}$ be enumerable disjoint sets of variables of type $\sigma$. 
The set of  \emph{ground} variables is $\Var^\iota$, 
the set of  \emph{higher-order} variables is $\hVar=\bigcup_{\sigma,\tau}\Var^{\sigma\lin\tau}$, the set of \emph{stable} variables is  $\sVar=\bigcup_{\sigma}\sVar^{\sigma}$
and the whole set of variables is $\Var=\Var^\iota\cup\hVar\cup \sVar$.
Letters  $\tt x^\sigma$ range over variables in $\Var^\sigma$,
letters $\tt y^\iota,z^\iota,\ldots$ range over variables in $\Var^\iota$,
letters $\tt f^{\sigma\lin\tau},g^{\sigma\lin\tau},\ldots$ range over variables in $\hVar$,
while $\tt \digamma^\sigma,{\digamma_1}^\sigma,{\digamma_2}^{\sigma},\dots$ range over stable 
variables, namely variables in $\sVar^{\sigma}$. 
Last, $\varkappa$ will denote any kind of variables. 
Latin letters ${\tt M,N,L,\ldots}$ range over terms.

A {\bf basis} $\Gamma$ is a finite list of variables in $\Var$. We denote with $\Gamma^*$ (resp $\Gamma^\iota$) a basis $\Gamma$ containing variables in $\sVar$ (resp. in $\Var^{\iota}$). We will denote with $\Gamma,\Delta$ the concatenation of 
two basis and with $\Gamma \cap \Delta$ the intersection of two basis, defined
in the expected way.
\begin{definition}
Typed {\em terms} of $\Sllambda$-calculus are defined by using a type assignment
proving judgements of the shape ${\Gamma} \vdash{\tt M:\sigma}$, in Table \ref{chapter1:fslPCF}.
\begin{table}[t]
\begin{center}
\begin{tabular}{|@{\hspace{1mm}}c@{\hspace{1mm}}|}
\hline
$\infer[\scriptsize\mbox{(z)}]{\vdash\num{0}:{\iota}}{\vspace{3mm}}
\quad
\infer[\scriptsize\mbox{(s)}]{\vdash\succ:{\iota\lin\iota}}{}
\quad
\infer[\scriptsize\mbox{(p)}]{\vdash\pred:{\iota\lin\iota}}{}
\quad
\infer[\scriptsize\mbox{($\lifz$)}]{{\Gamma , \Delta} \vdash \lif{{\tt M}}{{\tt L}}{{\tt R}}:{\iota}}
{{\Gamma\cap \Delta}= \emptyset\quad
{\Gamma}\vdash {\tt M}:{\iota}
\quad {\Delta}\vdash {\tt L}:{\iota}
\quad{\Delta}\vdash {\tt R}:{\iota}}
$
\\[3mm]
$
\infer[\scriptsize\mbox{(ex)}]{\Gamma,\varkappa_1^{\sigma_1},\varkappa_2^{\sigma_2},\Delta \vdash {\tt M} : \tau}{\Gamma,\varkappa_2^{\sigma_2},\varkappa_1^{\sigma_1},\Delta \vdash {\tt M} : \tau}
\qquad
\infer[\scriptsize\mbox{(gv)}]{{\tt x^\iota \vdash {\tt x}:{\iota}}}{}
\qquad
\infer[\scriptsize\mbox{(gw)}]{\Gamma , {\tt x}^\iota \vdash {\tt M} : \tau}{\Gamma \vdash {\tt M} : \tau}
\qquad
\infer[\scriptsize\mbox{(gc)}]{\Gamma , {\tt x}^\iota \vdash \tt M[{\tt x}/{\tt x_1},{\tt x_2}] : \tau}{\Gamma ,{\tt x_1}^\iota , {\tt x_2}^\iota \vdash {\tt M} : \tau}
$
\\[3mm]
$
\infer[\scriptsize\mbox{(hv)}]{{\tt {\tt f}^{\sigma\lin\tau}\vdash {\tt f}:{\sigma\lin\tau}}}{}
\qquad
\infer[\scriptsize\mbox{(ap)}]{\Gamma , \Delta \vdash {\tt M N:{\tau}}}
{\Gamma \cap \Delta = \emptyset \quad
{\Gamma}\vdash {\tt M}:{\sigma\lin\tau}\quad {\Delta}\vdash {\tt N}:{\sigma}}
\qquad
\infer[\scriptsize\mbox{($\lambda$)}]{{\Gamma} \vdash{\tt \lambda x^{\sigma}.M:{\sigma\lin\tau}}}
{{\Gamma,{\tt x}^{\sigma}} \vdash{\tt M}:{\tau}}
$
\\[3mm]
$
\infer[\scriptsize\mbox{(sv)}]{{\tt \digamma^\sigma \vdash {\digamma}:{\sigma}}}{}
\qquad
\infer[\scriptsize\mbox{(sc)}]{\Gamma , \digamma^\sigma \vdash {\tt M[\digamma/\digamma_1,\digamma_2}] : \tau}{\Gamma ,\digamma_1^\sigma , \digamma_2^\sigma \vdash {\tt M} : \tau}
\qquad
\infer[\scriptsize\mbox{(sw)}]{\Gamma , \digamma^\sigma \vdash {\tt M} : \tau}{\Gamma \vdash {\tt M} : \tau}
\qquad
\infer[\scriptsize\mbox{($\mu$)}]{\Gamma^\iota , \Delta^* \vdash{\tt \mu\digamma.M:{\sigma}}}
{\Gamma^\iota , \Delta^*,\digamma^\sigma \vdash{\tt M:{\sigma}}}
$
\\
\hline
\end{tabular}
\end{center}
\caption{Type assignment system for $\Sllambda$-calculus}\label{chapter1:fslPCF}
\end{table}
\end{definition}
Note that only higher-order variables are subject to syntactical constraints.
Except for the $\lifz$ construction typed by an additive rule
doing an implicit contraction, higher-order variables are treated linearly.
Ground and stable variables belong to distinct kinds only for sake of simplicity, their free use implies that
$\Sllambda$-calculus is not syntactically linear (in the sense of  \cite{paolini08ppdp}).

Free variables of terms are defined as expected.
A term ${\tt M}$ is {\em closed} if and only if $\FV({\tt M})=\emptyset$,
otherwise ${\tt M}$ is {\em open}.
Terms are considered up to $\alpha$-equivalence, 
namely a bound variable can be renamed provided no free variable is captured.
Moreover, ${\tt M}[{\num{n}}/{\tt y}]$, ${\tt M}[{\tt N}/{\tt f}]$ 
and ${\tt M}[{\tt N}/{\digamma}]$
denote the expected capture-free substitutions. 

\begin{definition}\label{reductionRULES}
We denote $\leadsto$ the firing (without any context-closure)
 of one of the following rules:
\begin{center}
 \begin{tabular}{l@{\hspace{9mm}}l@{\hspace{9mm}}l}
   $\tt (\lambda f^{\sigma\lin\tau}.M)N \leadsto_\beta M[N/f]$ &
$\tt (\lambda z^{\iota}.M)\num{n} \leadsto_{\iota} M[\num{n}/z]$ &
 ${\tt \mu \digamma.M} \leadsto_{\tt Y}{\tt M[\mu\digamma.M/\digamma]}$ \\
$\tt \pred\;(\succ\; \num{n}) \leadsto_{\delta} \num{n}$ &
$\tt \lif{\num{0}}{{\tt L}}{{\tt R}}\leadsto_\delta L$ &
 $\tt \lif{ \underline{\tt n\!+\!\!1}}{{\tt L}}{{\tt R}}\leadsto_\delta R$
\end{tabular}
\end{center}
We call redex each term or sub-term having the
shape of a left-hand side of rules defined above.
We denote $\redSl$ the contextual closure of $\leadsto$.
Moreover,  we denote $\redSl^*$ and $=_\Sl$ respectively,
the reflexive and transitive closure of $\redSl$
and the  reflexive, symmetric and transitive closure of $\redSl$.
\end{definition}

We remark that $ \leadsto_\beta$  formalises a 
call-by-name parameter passing in case of an higher-order argument.
On the other hand, $\leadsto_{\iota}$ formalises a call-by-value parameter passing, namely the reduction  can fire only when the argument is a numeral.
As done in \cite{berry85as}, it is easy to prove properties as 
subject-reduction, post-position
of $\delta$-rules in a sequence of reductions, the confluence
and a standardisation theorem.




\section{Categorical model of $\Sllambda$-calculus}
In this section we define the categorical model of $\Sllambda$-calculus and
we prove its soundness with respect to $=_\Sl$. We assume some familiarity
with the notions of monoidal categories, comonoids, comonads, adjunctions 
and monoidal functors. For an introduction, see \cite{maclane98book}. We begin by recalling the definition of Linear Category, given by Benton, 
Bierman, Hyland and de Paiva, which proposes a categorical notion of model for 
Intuitionistic Linear Logic.
\begin{definition}[Linear Category \cite{benton92csl}] A {\em Linear Category} 
$\mathcal L = \tuple{\category{L},!,\delta,\varepsilon,q,d,e}$ consists of
(1) a symmetric monoidal closed category 
$\tuple{\category{L},\otimes,\lin,\unitObj}$;
(2) a symmetric monoidal comonad called {\em exponential comonad}
$\tuple{!,\delta,\varepsilon,q_{A,B},q_{\unitObj}} : \category{L} \to \category{L}$, such that (i) 
for every free $!$-coalgebra $\tuple{!A,\delta_A}$ there are two 
distinguished monoidal natural transformations with components $d_A : !A \rightarrow !A \otimes !A$ and $e_A : !A \rightarrow \unitObj$ which form a commutative comonoid and are coalgebra morphisms;
(ii) whenever $f : \tuple{!A,\delta_A} \to \tuple{!B,\delta_B}$ is a coalgebra morphism  between free coalgebras, then it is also a comonoid morphism.
\end{definition}

A Linear Category provides a sound categorical model of Intuitionistic Linear 
Logic \cite{benton92csl}. A $\Sllambda$-category will be a Linear Category, 
thus every $\Sllambda$-Category is a model of Intuitionistic Linear Logic. 
However, it is necessary to augment it with other opportune features, 
in order to generate a sound categorical model of $\Sllambda$-calculus.

\subsection{Numerals}
First of all, we need a canonical object to interpret ground type $\iota$ and
opportune morphisms to interpret successor and predecessor. The following 
definition is an adaptation to monoidal categories of the definition of 
``simple object of numerals'' given in \cite{hyland00iandc}. 

\begin{definition}[Monoidal Object of Numerals] \index{Monoidal object of numerals}
Let $\category{C}$ be a symmetric monoidal category. Let $N$ be an object 
equipped with two morphisms $0 : \unitObj \to N$ and $succ : N \to N$. A 
{\em numeral} $n : \unitObj \to N$ is defined inductively as the map 
$0 : \unitObj \to N$ for the base case, while the map $n + 1 : \unitObj \to N$ 
is equal to $succ \comp n$. $N$ is said to be a {\em monoidal object of 
numerals} when it is also equipped with a morphism $pred : N \rightarrow N$ 
such that the following diagram commutes
$$
\xymatrix{
\unitObj \ar[r]^{n+1} \ar[rd]^n & N \ar[d]^{pred} \\
 & N
}
$$
\end{definition}
The definition above is very weak. It is in fact not required that given two 
numerals $m: \unitObj \to N$ and $n : \unitObj \to N$ with $n \neq m$ (viewed
as numbers), they are distinct morphisms in $\category{C}$. Moreover the definition 
given above
does not allow to represent neither recursive nor primitive recursive 
functions in $\category{C}$. An analogous situation is also present in
in the definition of simple object of numerals given in \cite{hyland00iandc}.

For sake of completeness, we now compare the above definition with the 
definition given in \cite{pare88sl}. It extends the notion of natural number 
object, which was specifically defined for cartesian categories in
\cite{roman89jpaa}, to any monoidal category.
\begin{definition}[\cite{pare88sl,mackie93acs}] \index{Natural number object} 
Let $\category{C}$ be a 
symmetric monoidal closed category. By a {\em natural number object} in 
$\category{C}$ we mean an object $N$ and two morphisms $0 : \unitObj \to N$ and 
$succ : N \to N$ such that, given any pair of morphisms $c : \unitObj \to A$
and $f : A \to A$ there is a unique $h : N \to A$ making the following diagrams
commute.
$$
\xymatrix{
\unitObj \ar[rr]^0 \ar[drr]^c && N \ar[d]^{h} \\
                              && A                   
}
\qquad\qquad
\xymatrix{
N \ar[rr]^{succ} \ar[d]^h &&  N \ar[d]^h \\
A \ar[rr]^f              && A
}
$$
\end{definition}
In \cite{pare88sl}, Par\'e and Rom\'an show that in any symmetric monoidal 
category $\category{C}$ with a natural number object, the theory of primitive 
recursive functions can be developed. This is done by considering the category 
of commutative co-monoids in $\category{C}$, which is 
cartesian \cite{pare88sl} and where the theory of natural number objects is 
well developed. In detail,
if $\tuple{C,d_C,e_C}$ and $\tuple{D,d_D,e_D}$ are two commutative co-monoids,
then its cartesian product is given by $\tuple{C \otimes D, d_C \otimes d_D,
e_C \otimes e_D}$, while the pairing and the projections are defined as
\begin{eqnarray*}
\pi_1 &\mbox{ is the composite of }&  C \otimes D \morph{\id{C} \otimes e_D} C \otimes \unitObj \morph{\unitRLaw} C \\
\pi_2 &\mbox{ is the composite of }& C \otimes D \morph{e_C \otimes \id{D}} \unitObj \otimes D \morph{\unitLLaw} D \\
\tuple{f,g} &\mbox{ is the composite of }& E \morph{d_E} E \otimes E \morph{f \otimes g} C \otimes D
\end{eqnarray*}
for $f : E \to C$ and $g : E \to D$. The terminal object is $\unitObj$.

More specifically in \cite{pare88sl} it is shown that if $N$ is a 
natural number 
object, then it is a commutative co-monoid, by taking the morphisms 
$w_N : N \to \unitObj$ and $c_N : N \to N \otimes N$ to be the unique 
morphisms making the following diagrams commute
$$
\xymatrix{
\unitObj \ar[r]^0 \ar[dr]^{\id{{\unitObj}}} & N \ar[d]^{w_N} \\
                                           & \unitObj                  
}
\qquad
\xymatrix{
N \ar[r]^{succ} \ar[d]^{w_N}     &  N \ar[d]^{w_N} \\
\unitObj \ar[r]^{\id{{\unitObj}}} & \unitObj
}
\qquad
\xymatrix{
\unitObj \cong \unitObj \otimes \unitObj \ar[r]^0 \ar[dr]^{0 \otimes 0} & 
N \ar[d]^{c_N} \\
             & N \otimes N                  
}
\qquad
\xymatrix{
N \ar[rr]^{succ} \ar[d]^{c_N}     &&  N \ar[d]^{c_N} \\
N \otimes N \ar[rr]^{succ \, \otimes \, succ} && N \otimes N
}
$$
Furthermore $0 : \unitObj \to N$ and $succ : N \to N$ are both comonoid 
morphisms.Thus, all numerals are co-monoid morphisms, and all primitive 
recursive 
functions can be represented, in the same way as they were represented in a 
Cartesian Category \cite{roman89jpaa}. Observe again that the above definition
of {\em natural number object} does not require that given two numerals 
$n : \unitObj \to N$ and $m : \unitObj \to N$ with $n \neq m$ 
(viewed as numbers) are distinct morphisms in $\category{C}$. But in 
\cite{pare88sl}, it has been shown that if this holds and if $\category{C}$ is 
monoidal closed, then $\category{C}$ is equivalent to the one-object 
one-morphism category (an analogous fact holds also for Cartesian Closed
Categories \cite{hyland00iandc}).

The following proposition is a corollary of the above statement.
\begin{proposition} Let $\category{C}$ be a symmetric monoidal closed category
with a natural number object $N$. Then $N$ is a monoidal object of numerals.
\end{proposition}
\begin{proof}
Let $h : N \to N \otimes N$ be the unique morphism making the following
diagrams commute (the pairing and projections in the category of commutative
comonoid of $\category{C}$ are defined above).
$$
\xymatrix{
\unitObj \ar[rr]^0 \ar[drr]^{\tuple{0,0}} && N \ar[d]^{h} \\
&& N \otimes N
}
\qquad
\xymatrix{
N \ar[rr]^{succ} \ar[d]^{h} && N \ar[d]^h \\
N \otimes N \ar[rr]^{\tuple{succ \comp \pi_1,\pi_1}} && N \otimes N
}
$$
Thus, a choice for $pred : N \to N$ could be the following
$$
pred \mbox{ is the composite of } N \morph{h} N \otimes N \morph{\pi_2} N
$$
It is not difficult to see that this choice of $pred$ satisfies usual 
equation.
\end{proof}

We now give a notion of natural number object in Linear Categories. We
observe that a monoidal object of numerals is too weak, in 
order to be a sound interpretation of the type $\iota$ of $\SlPCF$. The 
structure of monoidal object of numerals should be enriched to obtain an
{\em exponential object of numerals}; it will be a monoidal object of numerals 
with additional morphisms allowing to duplicate and weaken occurrences of them 
and whose other morphisms respects the comonoidal structure induced by the 
exponential co-monad. 

\begin{definition}[Exponential object of numerals] \index{Exponential object of numerals} Let 
  $\tuple{\category{L},!,\delta,\varepsilon,q,d,e}$ be a Linear Category. An
  {\em exponential object of numerals} is a $!$-coalgebra $\tuple{N,p}$ such 
  that
  \begin{enumerate}
  \item $N$ is a monoidal object of numerals.
  \item There exists two morphisms $w_N : N \rightarrow \unitObj$ and 
    $c_N : N \rightarrow N \otimes N$ which form a commutative co-monoid and 
    are such that
    \begin{enumerate}
    \item $0 : \unitObj \rightarrow N$ and $succ : N \rightarrow N$ are 
      both co-algebras and co-monoid morphisms.
    \item $p : N \rightarrow !N$ is a co-monoid morphism.
    \end{enumerate}
  \end{enumerate}
\end{definition}

\begin{proposition} \label{chapter3:exp_nat_obj}
Let $\tuple{\category{L},!,\delta,\varepsilon,q,d,e}$ be a Linear Category and
let $N$ be a natural number object such that $\tuple{N,p}$ is a $!$-coalgebra
satisfying
\begin{enumerate}
\item $p : N \rightarrow !N$ is a co-monoid morphism.
\item $0 : \unitObj \rightarrow N$ and $succ : N \rightarrow N$ are
  coalgebra morphisms
\end{enumerate} 
Then $\tuple{N,p}$ is an exponential object of numerals.
\end{proposition}

\subsection{$\Sllambda$-category}
We now introduce our categorical model, by defining the notion of 
$\Sllambda$-category whose morphisms will denote $\Sllambda$-terms. We 
introduce some notation on the symmetric monoidal closed category 
$\category{L}$ first. We let $\symLaw_{A,B} : A \otimes B \cong B \otimes A$ the
tensorial symmetric law. We denote with 
$\curry{-} : \homset{L}{C \otimes A}{B} \rightarrow \homset{L}{C}{A \lin B}$ 
the isomorphism induced by the canonical adjunction. When $C = A \lin B$,
we denote with $\eval : A \lin B \otimes A \to B$ the (unique!) morphism
such that $\curry{\eval} = \id{A \lin B}$. 

An 
$\Sllambda$-category is a Linear Category admitting an exponential object of 
numerals, together with a ``conditional-like'' morphism and a fix-point 
morphism for every object $B$ in the Kleisli category over the co-monad $!$, 
which is cartesian closed. This leads to the following definition.

\begin{definition}[$\Sllambda$ Category] \index{$\Sllambda$ category}
A $\Sllambda$ Category is a linear category $\mathcal{L} = \tuple{\category{L},!,\delta,\varepsilon,q,d,e}$ such that
  \begin{description}
  \item[Numerals.] $\category{L}$ admits and exponential object of numerals
    $\tuple{N,p}$.
  \item[Conditional Operator.] $\category{L}$ is cartesian and there exists a 
morphism $\ell if : N \otimes (N \product N) \rightarrow N$ such that, for all 
$f,g : \unitObj \rightarrow N$, the following diagram commutes
$$
\xymatrix{
\unitObj \cong \unitObj \otimes \unitObj \ar[rr]^{0 \otimes \tuple{f,g}} \ar[drr]_f & & N \otimes (N \product N)  
\ar[d]^{\ell if} & & \unitObj \cong \unitObj \otimes \unitObj\ar[ll]_{n+1 \otimes \tuple{f,g}} \ar[dll]^g \\
&& N &&
}
$$
\item[Fix-Point Operator.] The Kleisli category $\category{L}_!$ (which is Cartesian Closed) admits a fix-point operator $fix_B : !(!B \lin B) \rightarrow B$ for any object $B$. We remind that, by the Kleisli-construction, we have that the following diagram commutes.
$$
\xymatrix{
!(!B \lin B) \ar[d]^{fix_B}\ar[rr]^{d_{!B \lin B}} & & !(!B \lin B) \otimes !(!B \lin B) \ar[d]^{\varepsilon_{!B \lin B} \otimes (!fix_B \comp \delta_{!B \lin B})} \\
B & & (!B \lin B) \otimes !B \ar[ll]_{\eval}
}
$$
\end{description}
\end{definition}

\begin{definition}[Categorical $\Sllambda$-model] \label{chapter1:sllambda_cat_model} 
A categorical $\Sllambda$-model consists of
\begin{itemize}
\item A $\Sllambda$ Category $\tuple{\mathcal{L},N,p,c_N,w_N,\ell if, fix}$, where $\mathcal{L} = \tuple{\category{L},!,\delta,\varepsilon,q,e,d}$.
\item A mapping associating to every $\Sllambda$-type $\sigma$, an object $\den{\sigma}$ of $\category{L}$ such that $\den{\iota} = N$ and $\den{\sigma \lin \tau} = \den{\sigma} \lin \den{\tau}$.
\item Given a basis $\Gamma$ we define $\den{\Gamma}$ by induction as
$\den{\emptyset} = \unitObj$, $\den{{\tt x}^{\sigma} , \Delta} =
\den{\sigma} \otimes \den{\Delta}$ and $\den{\digamma^\sigma , \Delta} = 
!\den{\sigma} \otimes \den{\Delta}$.
Moreover, given a basis $\Gamma$ such that $\Gamma^\iota = {\tt x_1^\iota},\dots,
{\tt x_n^\iota}$ (resp. 
$\Gamma^* = \digamma_1^{\sigma_1}, \dots, \digamma_n^{\sigma_n}$) we denote with 
$p_\Gamma = {p \otimes \dots \otimes p}$ $n$-times 
(resp. $\delta_\Gamma = \delta_{\den{\sigma_1}} \otimes \dots \otimes \delta_{\den{\sigma_n}}$). \\
Given a term $\tt M$ such that $\Gamma \vdash {\tt M} : \sigma$ we associate it a morphism $\den{\Gamma \vdash {\tt M} : \sigma} : \den{\Gamma} \rightarrow \den{\sigma}$, such that
\begin{enumerate}
\item $\den{\vdash \num{0} : \iota} = 0$, $\den{\vdash \succ : \iota \lin \iota} = \curry{succ}$, 
$\den{\vdash \pred : \iota \lin \iota} = \curry{pred}$, $\den{{\tt x^\iota} \vdash {\tt x} : \iota} = \id{N}$
\item $\den{{\tt f}^{\sigma\lin\tau} \vdash {\tt f} : \sigma \lin \tau}^{\category{L}} = \id{\den{\sigma\lin\tau}}$, 
$\den{\digamma^{\sigma} \vdash \digamma : \sigma} = \varepsilon_{\den{\sigma}}$
\item $\den{\Gamma^\iota , \Delta^* \vdash \mu \digamma.{\tt M} : \sigma} = fix_{\den{\sigma}} \comp !\curry{\den{\Gamma^\iota , \Delta^* , \digamma^\sigma \vdash {\tt M} : \sigma}} \comp q \comp (p_\Gamma \otimes \delta_\Delta)$
\item $\den{\Gamma \vdash \lambda {\tt x}^{\sigma}.{\tt M} : \sigma \lin \tau} = \curry{\den{\Gamma , {\tt x}^{\sigma} \vdash {\tt M} : \tau}}$
\item $\den{\Gamma , \varkappa_1^{\sigma_1} , \varkappa_2^{\sigma_2} , \Delta \vdash {\tt M} : \tau} = \den{\Gamma , \varkappa_2^{\sigma_2} , \varkappa_1^{\sigma_1} , \Delta \vdash {\tt M} : \tau} \comp (\id{\den{\Gamma}^{\category{L}}} \otimes \symLaw_{\den{\sigma_1},\den{\sigma_2}} \otimes \id{\den{\Delta}})$
\item $\den{\Gamma , \Delta \vdash {\tt MN} : \tau} = \eval \comp (\den{\Gamma \vdash {\tt M}: \sigma \lin \tau} \otimes \den{\Delta \vdash {\tt N} : \sigma})$.
\item $\den{\Gamma , \Delta \vdash \lif{{\tt M}}{{\tt L}}{{\tt R}} : \iota} = 
\ell if \comp (\den{\Gamma \vdash {\tt M} : \iota} \otimes \tuple{\den{\Delta \vdash {\tt L} : \iota},\den{\Delta \vdash {\tt R} : \iota}})$.
\item $\den{\Gamma , {\tt x}^{\iota} \vdash {\tt M[x/x_1,x_2]} : \tau} = \den{\Gamma , {\tt x_1}^{\iota} , {\tt x_2}^{\iota} \vdash {\tt M} : \tau} \comp \id{\den{\Gamma}} \otimes c_N$
\item $\den{\Gamma , \digamma^{\sigma} \vdash {\tt M[\digamma/\digamma_1,\digamma_2]} : \tau} = \den{\Gamma , {\tt \digamma_1}^{\sigma} , {\tt \digamma_2}^{\sigma} \vdash {\tt M} : \tau} \comp \id{\den{\Gamma}} \otimes d_{\den{\sigma}}$
\item $\den{\Gamma , {\tt x}^{\iota} \vdash {\tt M} : \tau} = \den{\Gamma \vdash {\tt M} : \tau} \comp \id{\den{\Gamma}} \otimes w_N$
\item $\den{\Gamma , \digamma^{\sigma} \vdash {\tt M} : \tau} = \den{\Gamma \vdash {\tt M} : \tau} \comp \id{\den{\Gamma}} \otimes e_{\den{\sigma}}$
\end{enumerate}
\end{itemize}
\end{definition}
\section{Soundness}
The following theorem shows that the three kinds of syntactical substitutions 
are modelled by categorical composition of morphisms. Let us observe that the 
substitution of a ground or higher-order variable respectively with a numeral or 
a term is modelled directly with the composition in $\category{L}$, while the 
substitution of a stable variable with a term is modelled with the composition 
in the category of coalgebras.

\begin{theorem}[Semantical Substitution Lemma] \label{sub_lemma}$\;$
\begin{enumerate} 
\item Let $\tt M$ be such that 
$\Gamma , {\tt x}^\iota,\Delta \vdash {\tt M} : \sigma$. Then 
$\den{\Gamma,\Delta \vdash {\tt M[\num{n}/x]} : \sigma} = \den{\Gamma , {\tt x}^\iota,\Delta \vdash {\tt M} : \sigma} \comp (\id{\den{\Gamma}} \otimes n \otimes \id{\den{\Delta}})$.
\item Let ${\tt M,N}$ be such that $\Gamma, {\tt f}^\sigma \vdash {\tt M} : \tau$ and
$\Delta \vdash {\tt N} : \sigma$, with 
$\Gamma \cap \Delta = \emptyset$. Then \\
$\den{\Gamma , \Delta \vdash {\tt M[N/f]} : \tau} = \den{\Gamma , {\tt f}^\sigma \vdash {\tt M} : \tau} \comp (\id{\den{\Gamma}} \otimes \den{\Delta \vdash {\tt N} : \sigma})$.
\item
Let ${\tt M,N}$ be such that $\Gamma , \digamma^\sigma \vdash {\tt M} : \tau$ and $\Delta_1^\iota , \Delta_2^* \vdash {\tt N} : \sigma$, with $\Gamma \cap \Delta_1 \cap \Delta_2 = \emptyset$. Then \\ $\den{\Gamma,\Delta_1^\iota,\Delta_2^* \vdash {\tt M[N/\digamma]} : \tau} = \den{\Gamma , \digamma^\sigma \vdash {\tt M} : \tau} \comp (\id{\den{\Gamma}} \otimes (!\den{\Delta_1^\iota , \Delta_2^* \vdash {\tt N} : \sigma} \comp q \comp (p_{\Delta_1} \otimes \delta_{\Delta_2})))$.
\end{enumerate}
\end{theorem}
\begin{proof}
All the proofs follow by induction on the derivation of the typing judgements.
The key point is to show that the transformations induced by the typing rules 
are natural on the unchanged components of the sequent. More details can be 
found in \cite{piccolo09phd}.
\end{proof}
\begin{theorem}[Soundness] \label{chapter1:cat_soundness}
 Let $\tt M,N$ be such that $\Gamma \vdash {\tt M} : \sigma$ and $\Gamma\vdash {\tt N}:\sigma$. Then, 
$$\text{if }{\tt M =_\Sl N} \text{ then } \den{\Gamma \vdash {\tt M} : \sigma} = \den{\Gamma \vdash {\tt N} : \sigma}$$
\end{theorem}
\begin{proof} 
The proof is by induction on the derivation of $\tt M =_\Sl N$. We develop only
the case $\tt M =_\Sl N$ since $\tt M \leadsto_Y N$. Thus
$\tt M = \mu\digamma.M_1$ and $\tt N = M_1[\mu\digamma.M_1/\digamma]$.
To help with the notation, in the proofs we relax a bit the definition
of basis and we add types prefixed with a $!$ to the syntax of types. Thus,
given a basis $\Gamma = \varkappa_1^{\sigma_1} , \dots, \varkappa_n^{\sigma_n}$,
we denote with 
$!\Gamma = \varkappa_1^{!\sigma_1} , \dots \varkappa_n^{!\sigma_n}$ (where $!$ is
just a syntactical annotation which will be interpreted with the corresponding
categorical operator) and we adapt in the canonical way the interpretation
function on the so obtained types and basis.
First 
of all, if we let 
$f = \den{\Gamma^\iota,\Delta^*,\digamma^\sigma \vdash {\tt M}_1 : \sigma}$, 
let us observe that the following diagram commutes
$$
\xymatrix{
\den{\Gamma,\Delta} \ar[rr]^{\hspace{-10mm}(c_N \otimes \dots \otimes c_N) \otimes (d \otimes \dots \otimes d)} \ar[d]^{p_\Gamma \otimes \delta_\Delta}&& \den{\Gamma,\Gamma,\Delta,\Delta} \cong \den{\Gamma,\Delta,\Gamma,\Delta} \ar[d]^{p_\Gamma \otimes p_\Gamma \otimes \delta_\Delta \otimes \delta_\Delta} \ar[rr]^{\id{\den{\Gamma,\Delta}} \otimes p_\Gamma \otimes \delta_\Delta} && \den{\Gamma,\Delta} \otimes \den{!\Gamma,!\Delta} \ar[d]^{\id{\den{\Gamma,\Delta}} \otimes p_{!\Gamma} \otimes \delta_{!\Delta}}\\
\den{!\Gamma,!\Delta} \ar[rr]^{\hspace{-10mm}d\otimes \dots \otimes d} \ar[d]^{q}&& \den{!\Gamma,!\Gamma,!\Delta,!\Delta} \cong \den{!\Gamma,!\Delta,!\Gamma,!\Delta} \ar[d]^{q \otimes q} \ar[rr]^{\varepsilon_{\Gamma,\Delta} \otimes \delta_{!\Gamma,!\Delta}}&& \den{\Gamma,\Delta} \otimes \den{!!\Gamma,!!\Delta} \ar[d]^{\id{\den{\Gamma,\Delta}} \otimes (!q \comp q)}\\
!\den{\Gamma,\Delta} \ar[rr]^{d} \ar[d]^{!\curry{f}}&& !\den{\Gamma,\Delta}\otimes !\den{\Gamma,\Delta} \ar[d]^{!\curry{f} \otimes !\curry{f}}\ar[rr]^{\varepsilon_{\den{\Gamma,\Delta}} \otimes \delta_{\den{\Gamma,\Delta}}}&& \den{\Gamma,\Delta} \otimes !!\den{\Gamma,\Delta} \ar[d]^{\curry{f} \otimes !!\curry{f}}\\
!(!\den{\sigma} \lin \den{\sigma}) \ar[rr]^{\hspace{-5mm} d_{!\den{\sigma}\lin \den{\sigma}}} && !(!\den{\sigma} \lin \den{\sigma}) \otimes !(!\den{\sigma} \lin \den{\sigma}) \ar[rr]^{\varepsilon_{!\den{\sigma}\lin\den{\sigma}} \otimes \delta_{!\den{\sigma}\lin\den{\sigma}}}  && (!\den{\sigma} \lin \den{\sigma}) \otimes
!!(!\den{\sigma} \lin \den{\sigma})
}
$$
where the left square on the top commutes since $p$ and $\delta$ are co-monoid morphisms, the right square on the top commutes since $p$ and $\delta$ are co-algebras (observe that we used both commutative diagrams of the definition of co-algebra) and by bi-functoriality, the left square on the middle commutes since $d$ is a monoidal natural transformation, the right square on the middle commutes since $\delta$ and $\varepsilon$ are monoidal natural transformations, and finally the two squares on the bottom commutes respectively because being $!\curry{f}$ a co-algebra morphism between free co-algebra, it is also a co-monoid morphism, by naturality of $\varepsilon$ and $\delta$ and by bi-functoriality. Thus, we have,\\[3mm]
$
\begin{array}{l}
\den{\Gamma^\iota , \Delta^* \vdash {\tt M} : \sigma} = fix_{\den{\sigma}} \comp !\curry{f} \comp q \comp (p_\Gamma \otimes \delta_\Delta) \\
= \eval \comp (\varepsilon_{!\den{\sigma}\lin\den{\sigma}} \otimes (!fix_{\den{\sigma}} \comp \delta_{!\den{\sigma}\lin\den{\sigma}})) \comp d_{!\den{\sigma}\lin\den{\sigma}} \comp
!\curry{f} \comp q \comp (p_\Gamma \otimes \delta_\Delta) \\
= \eval \comp (\curry{f} \otimes (!fix_{\den{\sigma}} \comp !!\curry{f} \comp 
(!q \comp q) \comp (p_{!\Gamma} \otimes \delta_{!\Delta})) \comp (p_\Gamma \otimes \delta_\Delta))
\comp ((c_N \otimes \dots \otimes c_N) \otimes (d \otimes \dots \otimes d)) \\
= f \comp
\id{\den{\Gamma,\Delta}} \otimes (!\den{\Gamma^\iota,\Delta^* \vdash {\tt \mu\digamma.M} : \sigma} \comp q \comp (p_\Gamma \otimes \delta_\Delta)) 
\comp ((c_N \otimes \dots \otimes c_N) \otimes (d \otimes \dots \otimes d))\\
=\den{\Gamma^\iota,\Delta^*,\Gamma^\iota,\Delta^* \vdash {\tt N} : \sigma} \comp
 ((c_N \otimes \dots \otimes c_N) \otimes (d \otimes \dots \otimes d))
\end{array}
$\\[3mm]
where in the second line we used the fix-point law, in the third line we use the commutativity of the above diagram, in the fourth line we use the definition of interpretation, the naturality of $q$ and the fact that the category is monoidal closed. Finally, in the fifth line we use Theorem \ref{sub_lemma} point (3). Then we can conclude by definition of interpretation.
\end{proof} 

\section{Instances of $\Sllambda$-categories} 
\label{model-examples}
In this section, we show 
three interesting concrete 
instances of $\Sllambda$-category,
in the setting of Scott Domains
and Coherence Spaces. By means of the results proved in the 
previous section this three instances gives models in which the $\Sllambda$-calculus can be soundly interpreted.

\subsection{Scott Domains and strict continuous functions}
Let $\mathbf{StrictBcdom}$  (Strict Bounded Complete 
Domains) be the category obtained by taking as {\bf objects}
{\em $\omega$-algebraic bounded complete partial orders} (or 
{\em Scott domains}) and as {\bf morphisms} {\em strict continuous functions}, 
namely those continuous functions that map the bottom element of the source 
object to the bottom element of the target. This category is monoidal closed,
by taking the tensor product $A \otimes B$ to be the smash product 
$A \wedge B = \{\tuple{a,b} \mid a \in A \setminus\{\bot\} , b \in B \setminus\{\bot\}\} \cup \{\bot\}$ , the unit of the tensor product $\unitObj$ to be the 
Sierpinsky Domain $\{\bot,\top\}$ with $\bot \leq \top$ and the function space 
$A \lin B$ consisting of all strict maps between $A$ and $B$ under the 
point-wise order. Moreover if we take as exponential comonad $!$, the lifting
constructor $(-)_\bot$, we obtain a linear category; we remind that, given a 
Scott Domain $A$, the domain $A_\bot$ is obtained from $A$ by adding a new least
element below the bottom of $A$ (for more details see \cite{pravato99mscs}). 
Observe that the Kleisli category over the comonad $(-)_\bot$ is the usual 
category of Scott Domains and continuous functions.

We can prove that  $\mathbf{StrictBcdom}$ is a $\Sllambda$ Category, by taking 
$N$ to be the usual flat domain of natural numbers with the coalgebra 
$p : N \to N_\bot$ such that $p(n) = n$ for all $n \neq \bot$. $N$ is a 
commutative comonoid, by taking $w_N : N \to \unitObj$ be such that 
$w_N(n) = \top$ and $c_N : N \to N \otimes N$ be such that 
$c_N(n) = \tuple{n,n}$ for all $n \neq \bot$.  $\mathbf{StrictBcdom}$ is 
cartesian, by taking $A \product B$ to be the usual cartesian product of Scott
Domains. Thus we can define $\ell if : N \otimes (N \product N) \to N$ to be such
that $\ell if(c) = m_1$ if $c = \tuple{0,\tuple{m_1,m_2}}$, $\ell if(c) = m_2$ 
if $c = \tuple{n,\tuple{m_1,m_2}}$ and $n \neq 0$ and $\ell if(c) = \bot$ otherwise.
Finally, it follows easily by Knaster-Tarsky's Fix Point Theorem that the considered category admits fix point for every object.  This model is shown to be
adequate with respect to the operational semantics of $\SlPCF$ in 
\cite{piccolo09phd}.

\subsection{Coherence Spaces.}
A coherence space is a pair $X = \tuple{|X|,\coh_X}$, consisting of a finite or
countable set of tokens $|X|$ called {\em web} and a binary reflexive 
symmetric relation on $|X|$ called {\em coherence relation}. The set of 
{\em cliques} of $X$ is given by 
$\Cl(X) = \{x \subseteq |X| \mid a,b \in x \Rightarrow a \coh_X b\}$. This set
ordered by inclusion forms a Scott Domain whose set of finite elements is the set
$\Cl_{fin}(X)$ of finite cliques. Two cliques $x,y \in \Cl(X)$ are 
{\em compatible} when $x \cup y \in \Cl(X)$. 
A continuous function $f : \Cl(X) \rightarrow \Cl(Y)$ is {\em stable} when
it preserves intersections of compatible cliques. A stable function 
$f : \Cl(X) \rightarrow \Cl(Y)$ is {\em linear} when it is strict and preserves 
unions of compatibles cliques. Given a linear function $f : \Cl(X) \to \Cl(Y)$,
we denote its trace with $\tr(f) = \{(a,b) \mid b \in f(\{a\})\}$. We say that 
a linear function $f$ is less or equal than $g$ according to the 
{\em stable order} when $\tr(f) \subseteq \tr(g)$.

Let $\Coh$ be the category of coherence spaces as {\bf objects} and linear 
functions as {\bf morphisms}. Given a coherence space $X$, we define $!X$ 
to be the coherence space having as web the set $\Cl_{fin}(X)$ and as coherence
relation, the compatibility relation between cliques. It is possible to prove 
that $!$ is an exponential comonad, thus $\Coh$ is a Linear Category 
\cite{mellies03tr}. 
Observe that the Kleisli category over the comonad $!$ is the category of 
Coherence Spaces and Stable Functions

The model of $\SlPCF$ we define in \cite{paolini08ppdp} is based on this 
category, and can be obtained as follows. As in previous section, we take 
$N$ to be the infinite flat domain of natural numbers, and we define $w_N,c_N$ 
in an analogous way as before, as well as $\ell if$ and the fix point operator.
The $!$-coalgebra $p : N \to !N$ is such that $\tr(p) = \{(n,\{n\}) \mid 
n \in \mathbb{N}\} \cup \{(n,\{\emptyset\}) \mid n \in \mathbb{N}\}$.

\subsection{Scott Domains and Linear Functions}
A similar construction as the one presented above can be obtained also
in the Scott Domain setting. Let $\mathbf{LinBcdom}$ (Linear Bounded Complete 
Domains) be the category defined as follows. The objects are 
again {\em Scott Domains}. If $D$ is a Scott Domain, we write $D_0$ for its 
poset of finite elements; $D$ is obtained from $D_0$
by adding all suprema of directed subsets of $D_0$. The morphisms are {\em
linear maps}, i.e. functions which preserves all existing suprema 
($f : D_1 \to D_2$ is linear if for all bounded $X \subseteq D_1$ we have
$f(\bigsqcup X) = \bigsqcup f (X)$, reminding that 
$\bigsqcup \emptyset = \bot$). The tensor product $D_1 \otimes D_2$ classifies
maps $D_1 \times D_2 \to D$ linear in each argument, while the unit of the 
tensor
product $\unitObj$ is the Sierpinsky Domain $\{\bot,\top\}$ with 
$\bot \leq \top$; the linear function space $B \lin C$ consists of all linear 
functions from $B$ to $C$ ordered pointwise.
The cartesian product is the usual cartesian product between Scott Domain. The
exponential comonad can be described in terms of finite element. Given $D$, 
we let the set $(!D)_0$ to be the set obtained from $D_0$ by freely adding 
suprema of bounded finite subsets of $D_0$. We complete $(!D)_0$ with all 
directed limits, to obtain $!D$. The Kleisli category over the comonad is the
usual category of Scott domains and continuous maps.

We can prove that  $\mathbf{LinBcdom}$ is an $\Sllambda$ category by taking
$N$ to be the usual flat domain of natural numbers and defining 
$p : N \to !N$ as $p(\bot) = \bot$ and $p(n) = \sqcup \{\bot , n\}$, 
$w_N : N \to \unitObj$ as $w_N(\bot) = \bot$ and $w_N(n) = \top$, 
$c_N : N \to N \otimes N$ as $c_N(\bot) = \bot$ and $c_N(n) = \tuple{n,n}$.

\section{Towards Completeness}
\label{towards-completeness}
In the previous sections we have proved the soundness of our interpretation and we have shown some concrete examples. 
Now, a natural question is whether the $\Sllambda$-calculus is 
also \emph{complete}
with respect to the notion of $\Sllambda$-category we have
introduced so far or not. \\
The answer is negative. Indeed, the completeness of 
$\Sllambda$-calculus with respect to the notion of 
$\Sllambda$-model fails.
To understand why, let us consider the judgment 
$$
\Gamma,\Delta\vdash (\lambda {\tt x}^{\iota}. {\tt M}){\tt N}:\tau
$$
where $\Gamma \vdash \lambda {\tt x}.{\tt M}: \iota \lin \tau$
and $\Delta \vdash {\tt N} : \iota$. 
The interpretation of this judgment is 
$$
\den{\Gamma , \Delta \vdash {\tt (\lambda {\tt x}^{\iota}. {\tt M})N} : \tau} = \eval \comp (\den{\Gamma \vdash \lambda {\tt x}^{\iota}.{\tt M}: \iota \lin \tau} \otimes \den{\Delta \vdash {\tt N} : \sigma})
$$
The term ${\tt eval}$ above represents  the standard evaluation morphism
of the  symmetric monoidal closed category.
So, in particular it is easy to verify that the above interpretation
is equal to 
$$
\den{\Gamma , \Delta \vdash {\tt {\tt M}[{\tt N}/{\tt x}^{\iota}]} : \tau} 
$$
So, we clearly have:
$$
\den{\Gamma , \Delta \vdash {\tt (\lambda {\tt x}^{\iota}. {\tt M})N} : \tau} =
\den{\Gamma , \Delta \vdash {\tt {\tt M}[{\tt N}/{\tt x}^{\iota}]} : \tau} 
$$
Unfortunately, in the  $\Sllambda$-calculus we have 
\begin{equation}
\label{counterexample-Inequality}
 (\lambda {\tt x}^{\iota}. {\tt M}){\tt N}\neq_{\Sl} {\tt {\tt M}[{\tt N}/{\tt x}^{\iota}]}
\end{equation}
unless ${\tt N}$ is a numeral. So we have a counterexample to completeness.\\
In order to recover completeness, we can adopt different stategies.
First of all, note that the terms in the equation \ref{counterexample-Inequality} above
turn to be equivalent if we consider observational equivalence instead
of the equivalence induced by the reduction rules.
So we could consider $\Sllambda$-terms modulo observational equivalence. For such a system the completeness should hold. 
Unfortunately, this 
corresponds to study the notion of model of the 
programming language built over 
the calculus instead of studying a model of the calculus itself.
In our setting, this corresponds to study the notion of models for 
$\SlPCF$ instead of 
the $\Sllambda$-calculus but this is not our aim here.\\
A different perspective is to extend 
the term assignment system for ILL introduced in 
\cite{benton92csl} by means of operators for numerals,
conditional and fix points.
This can be done by extending the typed calculus by the rules in 
Figure \ref{chapter1:LinLambda} and by rules for conditional and fix points. The reduction rules of the obtained calculus should be
designed starting from the categorial equalities. So, for example
we obtain rules as\\[2mm]
$
\begin{array}{l}
    {\tt discard^{\iota} \ ({\tt succ }\ M)\ in\ N }\to
{\tt discard^{\iota} \  M\ in\ N  }\\
    {\tt discard^{\iota} \ 0\ in\ N }\to {\tt N  }\\
    {\tt copy^{\iota} \ ({\tt succ }\ M)\ as\ x^{\iota},y^{\iota}\ in\ N }\to
{\tt copy^{\iota} \  M\ as\ x^{\iota},y^{\iota}\ in\ N[{\tt succ\ x/x,succ\ y /y}]  }\\
    {\tt copy^{\iota} \ 0\ as\ x^{\iota},y^{\iota}\ in\ N }\to {\tt N[0/x,0/y]  }\\
\end{array}
$\\[2mm]
corresponding to the categorical equations deriving from the fact 
that zero and successor are comonoidal, as:\\[2mm]
$
  \begin{array}{l}
 {\tt discard\ {\tt (promote }^{\iota}\ M)\ in\ N }\to
{\tt discard^{\iota} \  M\ in\ N  }\\
    {\tt copy \ {\tt (promote^{\iota} }\ M)\ as\ x^{!\iota},y^{!\iota}\ in\ N }\to
{\tt copy^{\iota} \  M\ as\ x^{\iota},y^{\iota}\ in\ N[{\tt (promote^{\iota}\ x)/x, (promote^{\iota}\ y) /y}]  }\\
\end{array}
$\\[2mm]
corresponding to the categorical equations deriving from the fact that promotion on numerals is comonoidal
and as\\[2mm]
$
  \begin{array}{l}
{\tt derelict \ {\tt (promote^{\iota} }\ M)}\to {\tt M}\\
 {\tt promote\ {\tt (promote }^{\iota}\ M)\ as \ z\ in\ N[promote^{\iota}\ (derelict\ z)/z] }\to
{\tt promote (promote^{\iota} \  M)\ as\ z\ in\ N  }\\
\end{array}
$\\[2mm]
corresponding to the categorical equations deriving from the fact that
promotion on numerals is also a $!$-coalgebra.\\
Finally, another interesting possibility is to change the standard 
interpretation function. 
In particular, we could change the interpretation of 
$\lambda$-abstractions binding ground 
variables by adapting standard technics already studied for
the call by value $\lambda$-calculus.
\begin{table*}
\begin{center}
\begin{tabular}{|@{\hspace{1mm}}c@{\hspace{1mm}}|}
\hline\\[-2mm]
$
\infer[\scriptsize\mbox{(pr$_{\iota}$)}]{\Gamma\vdash {\tt promote^{\iota}(M)}:!{\iota}}{\Gamma\vdash {\tt M}:{\iota}}
\quad
\infer[\scriptsize\mbox{(ds$_{\iota}$)}]{\Gamma , \Delta \vdash {\tt discard^{\iota} \ M\ in\ N:{\sigma}}}
{
  {\Gamma}\vdash {\tt M}:{ \iota}& {\Delta}\vdash {\tt N}:{\sigma}}
\quad
\infer[\scriptsize\mbox{(cp$_{\iota}$)}]{\Gamma , \Delta \vdash {\tt copy^{\iota} \ M\ as\ x_1,x_2\ in\ N:{\sigma}}}
{
  {\Gamma}\vdash {\tt M}:{\iota }& {\Delta},{\tt x_1}^{\iota},
{\tt x_2}^{\iota}\vdash {\tt N}:{\sigma}}
$
\\
\\
\hline
\end{tabular}
\end{center}
\caption{Intuitionistic Linear typed calculus ILL}\label{chapter1:LinLambda}
\end{table*}

\section{Conclusion}
In this work we have introduced the notion of $\Sllambda$-category.
Such a notion provide a categorical model for $\Sllambda$-calculus introduced in \cite{paolini08ppdp}. We have shown that  $\Sllambda$-categories are sound for the interpretation of $\Sllambda$-terms. 
Moreover, we have shown three concrete model examples in the setting 
of Scott Domains and Coherence Spaces.\\
The  $\Sllambda$-calculus is not complete with respect to 
$\Sllambda$-category. In Section \ref{towards-completeness} we 
have sketched some approach in order to recover completeness.
We plan to explore these approaches in future developments.\\
The concrete  
denotational models presented in Section \ref{model-examples} can be 
useful in the
study of linear higher type computability \cite{longley02apal,paolini06iandc}.
In this setting one interesting research theme is the 
study of paradigmatic programming languages fitting models
founded on different higher type functionals.\\
On this matter, we have already obtained some preliminary results. 
In \cite{paolini08ppdp}
the interpretation of $\SlPCF$ into the category $\Coh$ is studied and a partial full
abstraction result is presented. 
In future works we plan to systematically
extend  $\SlPCF$ with suitable operators 
in order to establish definability results with respect to $\Coh$,
$\mathbf{StrictBcdom}$
and $\mathbf{LinBcdom}$. 


\bibliographystyle{eptcs}

\end{document}